\documentclass[10pt]{article}      
\usepackage{graphicx}
\usepackage{amscd,amsmath,amssymb,mathtools,cite,graphicx,color}
\usepackage{fancyhdr,bbold,authblk}
\usepackage{amsfonts,amsthm,eucal}
\usepackage{epstopdf}
\usepackage{epsfig}

\begin{document}

\title{The Definition of Entropy for Quantum Unstable Systems: A View-Point Based on the Properties of Gamow States}

\author{O. Civitarese$^1$, M. Gadella$^2$}

\maketitle

$^1$ Department of Physics, University of La Plata,  and IFLP-CONICET, 49 y 115. c.c.67, La Plata 1900, Argentina, osvaldo.civitarese@fisica.unlp.edu.ar 

$^2$ Departamento de F\'{\i}sica Te\'orica, At\'omica y \'Optica and IMUVA,  Universidad de Va\-lladolid, Paseo Bel\'en 7, 47011 Valladolid, Spain, manuelgadella1@gmail.com

\begin{abstract}

In this paper, we review the concept of entropy in connection with the description of quantum unstable systems. We revise the conventional definition of entropy due to Boltzmann and extend it  so as to include the presence of  complex-energy states. After introducing a generalized basis of states which includes resonances, and working with amplitudes instead of probabilities, we~found an expression for the entropy which exhibits real and imaginary components. We discuss the meaning of the imaginary part of the entropy on the basis of the similarities existing between thermal and time evolutions.  

\end{abstract}

\section{Introduction}
\label{1}

The definition of entropy and its interpretation in terms of the evolution to equilibrium of isolated systems was a crucial step in understanding  the link between mechanical and thermal features in classical mechanics \cite{REICHL}. The  notion of probability  
applies, both in classical phase-space as well as in quantum mechanics, and from this the connection between entropy and the number of degrees of freedom of a system has been established \cite{HUANG}. The main difference between classical and quantum mechanical counting of states is, of course, the existence of the exclusion principle (for fermions) and other symmetry restrictions (both for fermions and bosons) imposed to quantum states.~In~both cases, fermions and bosons, the definition of the probability assigned to a state remains valid. This~is not the case for states with complex energies, where the time evolution is non-oscillatory.~States~with complex energy, such as the Gamow states \cite{GAMOW}, are well described in the theory of scattering \cite{SCATTERING} and found as solutions of the analytical continuation of quantum relativistic and non-relativistic equations \cite{CG}. Several problems arise in dealing with these states, particularly their non-normalizability \mbox{\cite{BE,CGB}}. Most~of these difficulties are removed with the use of amplitudes, which are the solutions of the equations and/or with the corresponding propagators, instead of working with their modulus. A~suitable tool to work with Gamow states, in order to extract their thermodynamical
 information, is~the path integration.~In performing the path-integration we shall be dealing with amplitudes instead of probabilities, a~concept which cannot be applied to states with complex energy.

In the present article, we are going to show that a comprehensive scheme leading to the definition of entropy for resonances can be rigorously designed  by adopting path integration techniques. We~shall discuss this method as well as its application to a model for resonances which is analytically solvable.

The paper is organized as follows. In Section \ref{sec2} we revisit the conventional definition of entropy and relate it to time dependent operations, such as time inversion and time displacement.  Section \ref{sec3} deals with the identification of resonances in quantum physical systems and illustrate their time dependence. These properties are then shown to be found over solid  mathematical basis; e.g., we construct the decaying states in the framework of rigged Hilbert spaces \cite{BG}. Section \ref{sec4} is devoted to the notion of complex entropy and in Section \ref{sec5} we investigate the possible connection between our definition of complex entropy and the class of time operators \cite{MPC}. Our final remarks and conclusions are presented in Section \ref{sec6}.  

\section{Entropy and time Evolution}\label{sec2}

In the context of quantum mechanics in the Heisenberg picture, the time evolution of a system is governed by its Hamiltonian. Each operator  obeys the following equation of motion:

\begin{equation}\label{commu}
\left [ H, \cal{O} \right ]=-i \hbar \dot{\cal{O}}\,.
\end{equation}

In classical mechanics the commutator (\ref{commu}) is replaced by a Poisson bracket and the corresponding time evolution is determined by the classical Hamilton-Jacobi equations. From a perspective other than thermodynamics, the evolution of a system is determined by the extreme of its free energy. For~the moment we shall assume that the number of particles of the system is constant, this is why it makes sense to refer to the Helmholtz free energy $F=E-TS$. In such a circumstance,  the change of the entropy $S$ with respect to the energy, at constant temperature, is given by the equation

\begin{equation}\label{1}
\frac 1T=\frac{\partial S}{\partial\overline E}\,,
\end{equation}
where $\overline E$ is the mean value of the energy and  $T$ is the absolute temperature at which the
extreme of the free energy is reached (e.g., at the equilibrium). Though (\ref{1}) seems to belong to a class of equations of motion different from (\ref{commu}), the difference is only apparent, since both equations fix physical values at equilibrium. Then, we may establish a correspondence between a class of operators and the entropy, as the associated observable. We shall return to
this point later. 

It was Boltzmann who realized  that the number of degrees of freedom of a classical system is proportional to the logarithm of the number $\Omega$ of micro-states of the system, from where one derives the relation between the number of degrees of freedom and the  entropy, i.e., $S=-k\log \Omega$. The~way, in~which the entropy evolves as a physical system approaches to the equilibrium, is given by a celebrated theorem due to Boltzmann, the $\mathbb H$-theorem \cite{HUANG}. 
The $\mathbb H$-theorem states that if $P_r(t)$ is the probability that a system is in the state $r$ at time $t$ and if we define $\mathbb H:=\sum_r P_r(t)\log P_r(t)$, where the sum extends to all possible states of the system; then, $d\mathbb H/dt\le 0$.  The consequence is clear, since the entropy is given by $S=-k\mathbb H$, so that $dS/dt\ge 0$. The entropy monotonically increases with time until the system reaches the thermodynamic equilibrium. 

The same time evolution is expressed by means of the quantum evolution operator $e^{-itH}$, so~that if $O$ is the operator representing a given quantum observable at time $t=0$, the operator at time $t$ is given by $O(t)=e^{itH}Oe^{-itH}$ (Unless otherwise stated, we take $\hbar=1$ everywhere in the text.).

Therefore, it should exist a direct connection between both descriptions of the evolution to equilibrium.
However, from the time dependence of the observables of a system, one cannot always extract the direction of the evolution.   
The time reversal operation inverses the sense of time, so that it performs the operation $t\mapsto -t$. In classical mechanics, this means that the time reversal operation reverses momenta, velocities, etc, so that it reverses the velocities of the charges. This produces a change of the sign in the magnetic field, while leaves invariant the electric field. 

In quantum mechanics, the time reversal operation is represented by the action of an operator, $\cal{T}$, on the space of wave functions. According to Wigner \cite{W1}, time reversal is an operation such that the following operations performed sequentially give the identity:

\begin{eqnarray*}
\text{time displacement by } t \times \text{ time reversal }    \times  \text{ time displacement by } t \times \text{ time reversal}\,.
\end{eqnarray*}

The above operations result on the identity if
\begin{eqnarray*}
\text{time displacement by } t  \times \text{time reversal}  = \text{ time reversal} \times \text{time displacement by } -t\,.
\end{eqnarray*}

This point of view implies that the time reversal operator $\cal{T}$ has to be anti-linear in the sense that for any linear combination of states $\psi_i$ and two complex numbers $\lambda_i$, $i=1,2$, $\mathcal{T}(\lambda_1\psi_1+\lambda_2\psi_2)=\lambda_1^*\,\mathcal{T}(\psi_1)+\lambda^*_2\,\mathcal{T}(\psi_2)$, where the star denotes complex conjugation.  In fact, if $\psi(x,t)$ is the wave function for some quantum pure state at time $t$, we have that $\mathcal{T}\psi(x,t)=\psi^*(x,-t)$ \cite{GM}.
In addition, Wigner showed that in the construction of projective representations of the Poincar\'e group, extended  with time inversion and parity,  four independent choices exist for the time reversal operator. One~is the just mentioned operator $\cal{T}$ and the other three require a doubling of the representation space~\cite{W1,W2}. From a conceptual point of view, we are faced to a difficult question, namely: {\it{If equilibrium appears in a particular instant of the time evolution of a system and  is governed by a Hamiltonian, which is the operator that obeys Equation (\ref{commu}) and has the entropy given by the associated observable so that Equation (\ref{1}) is fulfilled?}} One~may also think that equilibrium is just a manifestation of the violation of the time-reversal symmetry, as shown by the behaviour of the entropy as a time dependent observable, as follows from the  $\mathbb H$-theorem.   

\section{Resonances in Quantum Systems}\label{sec3}

As is well known, unstable quantum states are very frequent in Nature.   They are characterized by two parameters:  $E_R$ and  $\Gamma$, which are the real and imaginary parts of the energy, respectively. The~quantity $\Gamma$ is the inverse of the state half-life. Usually, one may consider that unstable quantum states are produced by the capture of a particle by a center of forces and its subsequent decay, a~situation which is conveniently described by quantum scattering. The process of capture is often ignored as one is mainly concerned with the process of decay \cite{BG,CG}. They are detected experimentally by the presence of some scattering features, such as a sharp bump in the cross section or a sudden change in the value of phase shifts. Due to this fact, unstable quantum states are usually called resonances. We shall use this denomination hereafter.

After this characterization of resonances  in the context of scattering theory, they can be identified with poles in the analytic continuation of the $S$-matrix, provided that some smooth conditions be satisfied  \cite{BOHM}. If this analytic continuation is performed in the energy representation the $S$ matrix becomes a function of a complex energy defined on a two sheeted Riemann surface \cite{BOHM}. Resonances appear as pairs of complex conjugate poles located on the second Riemann sheet at the points \mbox{$z_R=E_R\pm i\Gamma/2$}, where $E_R>0$ is the resonance energy and $\Gamma>0$ the inverse of the half life, as~said~before. 

The description of a quantum scattering process  requires of two Hamiltonians. One is the free Hamiltonian $H_0$ that gives the free evolution of states. The other is a total Hamiltonian $H=H_0+ V$, where $V$ is the potential which produces the scattering. In the case of having resonances due to  scattering, the potential $V$ determines the forces that produce the capture and the later decay of the resonant particle.

A particularly interesting model for quantum resonances is the Friedrichs model \cite{F,HM,GP}. In~the simplest formulation of the Friedrichs model a bound state interacts with an external field. As the result of this interaction, the bound state becomes unstable and, therefore, it is interpreted as a resonance. In~the language of the Hamiltonian pair $\{H_0,H\}$, we have that 

\begin{equation}\label{2}
H_0= \omega_0\, a^\dagger\,a+\int_0^\infty \omega \,b^\dagger_\omega\,b_\omega\,d\omega\,, \quad V=\lambda \int_0^\infty f(\omega) [a^\dagger\,b_\omega+a\,b^\dagger_\omega]\,d\omega\,.
\end{equation}

We see that $H_0$ is the sum of two terms. In the former, $a^\dagger$ and $a$ are, respectively, the creation and annihilation of a bound state of energy $\omega_0>0$. The integral term in $H_0$ is the simplest representation of a field in the energy representation, where $b^\dagger_\omega$ and $b_\omega$ are, respectively, the creation and annihilation operators of a state in the continuum with energy $\omega>0$. Thus, $H_0$ has a  non-degenerated continuous spectrum, $[0,\infty)$, plus a discrete eigenvalue $\omega_0>0$ imbedded in the continuum.  The potential $V$ intertwines discrete and continuous spectrum, where $f(\omega)$ is a regular function called the form factor and $\lambda$ a real coupling constant. When the interaction $V$ is switched on, the bound state becomes a resonance with complex energy given by

\begin{equation}\label{eq3}
z_R=   E_R-i\Gamma/2\,.
\end{equation} 

Observations on the resonance decay show that the decay rate is approximately given by $e^{-t\Gamma/2}$. Now, if any bound state is represented by a square integrable wave function, is this the same for unstable quantum states (resonances)? Let us assume that a resonance state is represented by a vector state $\psi$. The survival amplitude is defined as $\langle\psi|e^{-itH}|\psi\rangle$ and the survival probability, $P(t)$, as the modulus square of the survival amplitude, i.e., $P(t)=|\langle\psi|e^{-itH}|\psi\rangle|^2$. If $\psi$ is to represent a resonance state, we expect that $P(t)\approx e^{-t\Gamma}$ for all values of $t$. 

However, this is not the case. In general, one may prove that there exist states $\psi$ for which $P(t)$ is approximately $e^{-t\Gamma}$ for most of observational values of time. These states may serve as resonance states. However, simple theoretical considerations show that deviations from the exponential decay law must exist for small and large values of time. These deviations are essential, i.e., they are a consequence of quantum theory, in particular of the semi-boundedness of the Hamiltonian, and not the product of noise or other interactions \cite{FGR}. There exists some experimental evidence on the existence of such deviations
\cite{exp1,exp2}. 

As a matter of fact,  $P(t)=|\langle\psi|e^{-itH}|\psi\rangle|^2$ shows a similar behaviour at very small times for all scattering states $\psi$. A simple calculation shows that $P'(0)=0$, where $P'(t)$ is the derivative of $P(t)$ with respect to $t$. This has a subtle consequence known as the Zeno effect: sequential (repeated)  measurements of the decay probability at very short intervals of time may prevent a decaying system to decay \cite{MISRA}. (This is the origin of the deviations of the purely exponential decay law for very short times, since if $P(t)=e^{-t\Gamma}$, then $P'(0)=-\Gamma\ne 0$.)

Nevertheless, a wide range of experiments on decaying systems show that the exponential decay is a good approximation  for most purposes. Then, the consideration of states that have a purely exponential decay should be in order.  These states can be rigorously constructed as eigenvectors, $\psi^D$, of $H$ with eigenvalue $z_R$ as in \eqref{eq3}, $H\psi^D=z_R\,\psi^D$. Each of the eigenvectors $\psi^D$ is called a {\it decaying Gamow vector} and has the property that $e^{-itH}\psi^D=e^{-iE_R t}\,e^{-\Gamma t/2}\psi^D$, i.e., it decays exponentially as $t\longmapsto\infty$. Since $\psi^D$ is an eigenvector of a self adjoint Hamiltonian with complex eigenvalue, then $\psi^D$ cannot belong to the Hilbert space where $H$ is densely defined and self-adjoint. Instead, it belongs to the dual $\Phi^\times$ of a rigged Hilbert space $\Phi\subset\mathcal H\subset\Phi^\times$. 

If a normalizable vector $\psi$ is taken to represent a resonance state, one may write $\psi=\psi^D+\psi^B$, where $\psi^B$ accounts for the deviations from exponential law for very short and very large times~\mbox{\cite{FGR,CG}}. Except for these two regimes of time, $\psi^D$ is a good approximation for $\psi$. However, as $\psi^D$ is not normalizable in the usual sense, one finds methodological difficulties to define mean values of observables on $\psi^D$ \cite{BE,CGB}. These difficulties will re-emerge as one attempts to assign a value to the entropy for quantum decaying systems.

\section{Complex Entropy}\label{sec4}

It should be clearly stated that any quantum unstable system should obey the laws of thermodynamics. The point is that a precise formulation of these laws for quantum decaying states has not been formulated yet,  up to our knowledge. Also, quantum statistical mechanics should extend its scope so as to embrace these kind of systems.

Based on this idea, one may ask for a suitable definition of the entropy for quantum unstable systems. At least three approaches have been proposed. A first approach was proposed by the Brussels group, it relies on the construction of an {\it entropy operator}, defined as a  monotonic function of the time operator \cite{MPC, KOPP}, which can be rigorously defined from a mathematical point of view under reasonable physical properties, see \cite{SS}. This way has not been fully explored yet.

A second approach was suggested by Kobayashi and Shimbori \cite{SH,KO}. There the entropy for a quantum unstable state, described  by a pole of the form $z_R=E_R-i\Gamma/2$, is a sum of a contribution of the entropy of the real part, $E_R$, and a contribution from the imaginary part $\Gamma$, so that $S=S(E_R)+S(\Gamma)$~\cite{KO}. This keeps the entropy as a real function of the resonance pole. In fact, real and imaginary parts of the complex resonance energy $z_R$ are treated as if they were two independent systems. In this picture, decaying processes transfer entropy  from $S(\Gamma)$ to $S=S(E_R)$ and the rate of this transference depends on time. Each part has its own temperature, which suggests a notion of complex temperature.

We advocate a third approach, which does not make use of the entropy operator and avoids any possible reference to complex temperatures. Following a suggestion in \cite{SH}, we assume that quantum decaying states are in thermodynamic equilibrium, provided that the half life be sufficiently large, or~equivalently, that~the imaginary part of its energy, $\Gamma$, be sufficiently small. 

Then,  in order to give a definition of the entropy for quantum unstable states, we need a universal model of resonances for which mathematical operations could be performed as much as possible. This~is given by the Friedrichs model described in the previous section. In the Friedrichs model, resonances are produced after the interaction of a discrete bound state with a continuum of states with a much larger degeneracy, so that it may be taken as a good example of a situation amenable to  a statistical description based on the canonical ensemble representation, where the continuum is playing the role of the environment interacting with the isolated discrete state. For simplicity, we may consider a Friedrichs model with one resonance only, although more complicated models could be used for the same purpose \cite{GP}. As is well known, the canonical entropy is given by the formula:

\begin{equation}\label{4}
S=k\left( 1-\beta\,\frac{\partial}{\partial\beta} \right) \log Z\,,
\end{equation}
where $Z={\rm Tr}\,e^{-\beta H}$ is the partition function corresponding to the total hamiltonian $H$ and $\beta=1/(kT)$, where $T$ is the absolute temperature and $k$ the Boltzmann constant.  

In order to evaluate $Z={\rm Tr}\,e^{-\beta H}$, it seems reasonable to use a generalized basis of vectors which includes the Gamow state $\psi^D$. This is given by $\psi^D$ and the so called generalized outgoing eigenvectors of the total Hamiltonian $\{|\omega^+\rangle\}$, with $H|\omega^+\rangle=\omega\,|\omega^+\rangle$, for all $\omega\ge 0$. Then, the partition function would have taken the following form:

\begin{equation}\label{5}
Z={\rm Tr}\,e^{-\beta H}= \langle \psi^D|e^{-\beta H}|\psi^D\rangle+ \int_0^\infty \langle\omega^+|e^{-\beta H}|\omega^+\rangle\,d\omega\,.
\end{equation}

However, this formula is not computable, as brackets of the form $ \langle \psi^D|\psi^D\rangle$ or $\langle\omega^+|\omega^+\rangle$ are not well defined \cite{GP,CGB,AP}. 

Then, we have to circumvent this problem by using a different technique based on the use of path integrals to calculate partition functions as introduced by Feynman and Hibbs \cite{FH}. In our approach, we have adopted path integration in order to write the partition function using a basis of coherent states. Thus, we construct coherent states in the following form: creation, $A_{\rm IN}^\dagger$, and annihilation, $A_{\rm OUT}$, operators for the Gamow state $\psi^D$ may be constructed for the second quantized Friedrichs model as described in \cite{GP}. Then, for any complex number $\alpha$, we define the coherent state $|\alpha\rangle$ as:

\begin{equation}\label{6}
|\alpha\rangle = \exp [\alpha\, A_{\rm IN}^\dagger-\alpha^*\,A_{\rm OUT}] \,|0\rangle\,,
\end{equation}
where $|0\rangle$ is the vacuum state and the star denotes complex conjugation. Then, an evaluation of the partition function,
although somehow cumbersome, can be done. Details are given in \cite{I,II,III}. We~arrive to the following result for the entropy of a Gamow state with complex energy $z_R=E_R-i\Gamma/2$:

\begin{equation}\label{7}
S=k\left[ 1-\ln \left( \beta\sqrt{E_R^2+\frac{\Gamma^2}{4}} \right) -i\arctan\left( \frac{\Gamma}{2E_R} \right) \right]\,.
\end{equation}

The result for $S$ is complex and this fact requires of some comments. Firstly, the method used to obtain the above formula is a straightforward  generalization of a similar method, which uses path-integration and coherent states, developed to obtain an approximation to the entropy of the harmonic oscillator \cite{III} avoiding the use of probabilities.
For the case of the harmonic oscillator it yields $S\approx k(1-\log(\beta\hbar\omega))$. Note that this is exactly the result that we obtain in the limit $\Gamma\mapsto 0$ and $E_R=\hbar\omega$. 
Secondly, since quantum resonances have complex energies with a different physical interpretation of real and imaginary parts, it is not a surprise that the same situation arises for the entropy. The resonance in the Friedrichs model is produced by the interaction of the bound state with the external field that plays the role of an external bath \cite{AGKPP}. With this idea in mind, one may interpret the real part of \eqref{7} as the system entropy and its imaginary part as the entropy transferred from the resonance to the external bath. 

There is another approach, described in \cite{III}, which leads to a complex entropy for an unstable quantum state. It is based on the fact that the total Hamiltonian has the form $H=z_R\, A^\dagger_{\rm IN}\,A_{\rm OUT}$, plus~a much smaller background term which is neglected.  Then, by using the property that the trace is invariant under cyclic permutations and formulas like 

\begin{equation}\label{8}
[H,A^\dagger_{\rm IN}(\tau)]=\frac{\partial }{\partial\tau}\, A^\dagger_{\rm IN}(\tau)=z_R\,A^\dagger_{\rm IN}\,;\,\, [H,A_{\rm OUT}(\tau)]=\frac{\partial}{\partial \tau}\,A_{\rm OUT} =-z_R\,A_{\rm OUT}\,,
\end{equation}
and for operators of the form $O(\tau)=e^{\tau H}\,O\,e^{-\tau H}$ with $\tau=\beta$, we obtain the desired result, of which~\eqref{7} could be considered as a reasonable approximation. Note that the definition for $O(\tau)$ has a great similarity with the definition for the time evolution of an operator, as suggested before (see Section \ref{sec2}).

\section{Time-Temperature Plane}\label{sec5}

Let us consider a quantum observable $O$ and define $O(\tau)$ as in the previous section, right after formula \eqref{8}, where  $\tau=\beta=1/(kT)$, being $T$  the absolute temperature. $O(\tau)$ denotes the thermal evolution of the observable. On the other hand, if we consider the time evolution of the observable $O$ under a Hamiltonian $H$, we have that $\tau=-it$. The transformation from the first to the second is sometimes called the Wick rotation \cite{REICHL}.  

This suggests the possibility of a description of time evolution for non-equilibrium systems using the dependence on both variables time and inverse temperature. The picture would be a complex plane in which the real part is given by the inverse of the temperature and the imaginary part by time. Similar notions have been applied to introduce dual-thermal degrees of freedom and close-path integrals \cite{keldish}.

Time operators have been defined for different purposes and different contexts \cite{AS,SA}.  For~instance, assume that $H$ is a densely defined Hamiltonian on a given Hilbert space $\mathcal H$. This~system has an internal time operator $\mathbb T$ if for any density operator $\rho$ in a domain dense in the space of Hilbert-Schmidt operators on $\mathcal H$, we have that $e^{-itH}\,\mathbb T\, e^{itH}\,\rho=(\mathbb T+tI)\rho$, for any $t$ real, where $I$ is the identity. Not any Hamiltonian system may have an internal time operator \cite{FZV}. 

One of the interesting aspects of time operators is the possibility  of constructing Liapunov quantum variables, i.e., variables monotonic on time. One may understand the role of such variables as indicators of the approach to equilibrium for complex systems, particularly for a definition of an entropy operator. Here, we refer to a construction proposed in \cite{MPC} and valid for a large class of situations. A necessary condition for the 
use of the procedure outlined in \cite{MPC} 
is that the Hamiltonian be unbounded and have an absolutely continuous spectrum, usually the half line $[0,\infty)$, 
if this condition is fulfilled $H$ is said to be semi-bounded.

The idea of a time operator emerges from the comparison between the position-momentum and energy-time uncertainty relations. However, and due to the semi-boundedness of the Hamiltonian in non-relativistic quantum mechanics, a commutation relation of the type $[H,\mathbb T]=iI$ cannot hold. In~any case, if the Hamiltonian is semi-bounded with absolutely continuous spectrum, its corresponding Liouvillian $L=H\otimes I-I\otimes H$ has a continuous spectrum that covers the whole real axis. In this case, it may be possible to define a time operator $\mathbb T$ as the conjugate of the Liouvillian operator, $[L,\mathbb T]=-iI$. Note that these operators have to be defined on $\mathcal H\otimes \mathcal H$, i.e., the space of Hilbert-Schmidt operators on~$\mathcal H$. 

If this were the case, one may define the entropy as some monotonic function of the time operator, as done in \cite{MPC}, i.e.,  $S=f(\mathbb T)$. Attempts to define a time operator, and hence an entropy operator for unstable decaying systems are on the course. 

\section{Final Remarks}\label{sec6}

In classical mechanics, the approach to equilibrium is a manifestation of the time reversal symmetry breaking. This is formulated via the Boltzmann  $\mathcal H$ theorem according to which the entropy monotonically increases up to a critical point, usually a maximum, at equilibrium. In classical electrodynamics, the retarded solutions of the Maxwell equations are privileged over the advanced solutions, thus showing a time asymmetry. 

In quantum mechanics, the decay of unstable systems such as quantum resonances  gives a sense of  time reversal symmetry breaking. One finds a need for a proper formulation describing this situation in a similar context as in classical mechanics, whenever possible. Then, it seems necessary to define the notion of entropy for quantum decaying systems. 

We have introduced an idea toward a proper definition of this entropy based on the use of Gamow states as state vectors for resonances. However, a naive presentation using standard tools of quantum mechanics yields to inconsistencies due to the ill definition of some formal averages. We have shown that the use of path integration over coherent states, which have been constructed with the help of creation and annihilation operators of Gamow vectors, gives a reasonable outcome. The resulting entropy is complex, with an imaginary part which gives an account for the interactions of decaying states with their surroundings. 

We have discussed the formal similarities between thermal and time evolution of states. Concerning quantum decaying 
states, we have introduced a representation, \eqref{8}, that gives the thermal evolution of the creation and annihilation operators for the Gamow states. These are differential equations that admit the following solutions:
\begin{equation}\label{9}
A_{\rm IN}^\dagger(\tau) =e^{\tau z_R}\, A_{\rm IN}^\dagger\,, \quad {\rm and} \quad A_{\rm OUT}(\tau)=e^{-\tau z_R}\,A_{\rm OUT}\,.
\end{equation}

Identities \eqref{9} are useful in order to obtain an expression for the complex entropy valid for the quantum unstable state created by $A_{\rm IN}^\dagger$ and annihilated by $A_{\rm OUT}$. Here, we choose $\tau=\beta=1/(kT)$,  as~given in \cite{III}. In this case, and since the resonance energy $E_R$ is taken to be positive, the highest the temperature, the smaller $A_{\rm IN}^\dagger(\tau)$ and the larger $A_{\rm OUT}(\tau)$. 

A completely different interpretation comes when $\tau=-it$, i.e., when we consider the time evolution of Gamow states. 
Now, $A_{\rm IN}^\dagger(t)=e^{-itE_R}\,e^{-\Gamma t/2}\, A_{\rm IN}^\dagger$ and $A_{\rm OUT}(t)=e^{itE_R}\,e^{\Gamma t/2}\,A_{\rm OUT}$, so that the creation operator for a Gamow state decays with time while the annihilation operator grows with time. This is called the Wick rotation. 

A future perspective could be a definition of the entropy operator as a function of the time operator. This is defined on an algebras of observables where  Gamow states play a role as functionals over this algebra. A prototype of this algebraic model has been constructed \cite{CGIL} and the investigation is on the course.

\section*{Acknowledgements}

 Partial financial support in acknowledged to the National Research Council of Argentina (CONICET) PIP 616 , to the Agencia Nacional de Promocion Cientifica y Tecnica de Argentina (ANPCYT), the Project MTM2014-57129-C2-1-P of the Spanish Ministerio de Econom\'ia y Competitividad and the Project VA057U16, awarded by the Junta de Castilla y Le\'on (Spain).

\end{document}